\documentstyle[mprocl]{article}

\def\Journal#1#2#3#4{{#1} {\bf #2}, #3 (#4)}

\def\be{\begin{equation}}
\def\ee{\end{equation}}
\def\bea{\begin{eqnarray}}
\def\eea{\end{eqnarray}}
\begin{document}

\title{TOWARD THE NEW GRAVITATIONAL NONCOMMUTATIVE MECHANICS AND
STATISTICAL MECHANICS OF QUANTUM BLACK HOLES}

\author{ P. O. MAZUR }

\address{Department of Physics and Astronomy
\\ University of South Carolina \\ Columbia, SC 29208, USA}


\maketitle\abstracts{This is a short account
of our work on the statistical mechanics
of `cold' quantum black holes in the constituent model of a black
hole~\cite{pom1}~\cite{pom2}~\cite{gorpom}.
A quantum Schwarzschild black hole
consists of {\em gravitational atoms} of Planckian mass
scale~\cite{pom1}~\cite{pom2}~\cite{gorpom}.
The gapless bosonic collective excitations of a bound state of $N$
{\em gravitational atoms} dominate the thermodynamics
of a cold quantum black hole.
It turns out that it is only in the limit of large $N$, with the
observable mean values of the gravitational mass-energy kept fixed,
that we recover the Bekenstein thermodynamics~\cite{bek}
of spin-zero black holes.
This is also the limit of the Boltzmann statistics.}

\section{On the Statistical Mechanics
of Quantum Schwarzschild Black Holes}

We have shown that a gravitating mass $M$
in thermal equilibrium behaves
statistically like a system
of some number $N$ of harmonic oscillators
whose zero-point energy ${\epsilon\over 2}$
depends on $N$ universally
in such a way that
$N{\epsilon}^2\sim {{hc^5}\over G}$ ,
where $G$ is the Newton constant,
$c$ is the velocity of light
in vacuum, and $h$ is the Planck constant.
The large number $N$ is also
the number of weakly interacting
gravitational atoms~\cite{pom2}~\cite{gorpom}~\cite{pom6}
which are the constituents of a spin-zero black hole.
The sum over all oscillators
of the squares of zero-point energies
is fixed and independent of
the number of those oscillators.
%It is well known that
%the classical gravitating systems behave
%in the way foreign to statistical quantum mechanics.
%The negative specific heat
%of those systems and the phenomenon
%of a gravitational collapse
%are different facets of the same reality.
The enigmatic Bekenstein entropy
of black holes was not yet derived
on the basis of microscopic theory.
Our work may be considered a
first step in direction of presenting
such a
basis~\cite{pom1}~\cite{pom2}~\cite{pom3}~\cite{pom5}~\cite{pom6}.
The problem with all present
approaches to this problem has been the
silent assumption that
the total entropy of gravitating systems
(black holes) {\it must} be given
by the Bekenstein formula~\cite{bek}
$S_{bh} = 4k{\pi}M^2$ ,
where the mass $M$ is in Planck units.
The postulate of gravitational constituents (gravitational
atoms) and gravitational oscillators (quanta)
leads to Bekenstein formula only after a part of mass-energy
fluctuations is neglected~\cite{pom2}~\cite{gorpom}.
The most unusual character of the gravitational mass-energy
oscillators (quanta) is that they somehow manage, via the quadratic
sum rule defining the Newton constant $G$,
$\sum_i {\epsilon_i}^2 = b{hc^5 \over G}$ ,
to reduce their zero-point energy when
the number $N$ of gravitational atoms grows~\cite{pom2}.
The formula: $M^2 ={{\mu^2}\over 2\pi}N$ ,
where ${\mu}^2 = {hc\over G}$,
was derived long time ago (1986)
by the present author.
This also means that a cold
large gravitational mass
$M\sim \mu{\sqrt N}$ consists of $N$
{\it constituents}.
The physical meaning of the
`phenomenological' entropy of Bekenstein~\cite{bek} is that
it is the measure of the number $N$ of {\it constituents}
making up a very `cold' large body. The zero-point energy $\epsilon_i$
of {\it gravitational quanta} for a very large `cold' mass is of an
order of the Hawking thermal energy of quanta~\cite{hawk},
${\epsilon_i}\sim {{\mu^2}\over (4{\pi})^2 M}$.
The more massive
is a gravitating mass the softer are the {\it gravitational quanta}.
The number $N$ of {\it constituent gravitational atoms} of a
given spin-zero quantum Schwarzschild black hole determine
the energy of quasi-thermal quanta and the Bekenstein-Hawking
entropy, $kT_{bh}\sim {\mu\over \sqrt{N}}$, $S_{bh}\sim kN$, but it
is valid only in the particular limit when the interference terms are
neglected~\cite{pom2}. Otherwise, as usual
with oscillators, there are two sources of statistical fluctuations
of mass-energy corresponding to the corpuscular and wave
aspect of quanta~\cite{pom2}~\cite{gorpom}. We calculate the energy
fluctuations of the system\footnote{Two examples of gravitating
systems are discussed in~\cite{pom2}\cite{gorpom}
simultaneously: a spin-zero quantum black hole and the whole
non-rotating Universe. The present upper bound on cosmological constant
$\lambda$ in Planck units,
regarded as the vacuum zero-point energy density,
was used to estimate the lower bound on
the total number $N_U\sim 10^{123}$ of gravitational atoms
(and gravitational oscillators) in the Universe.}
considered in~\cite{pom2}
in terms of the mean energy
$\overline{E}$, where
$\overline{E} = {E_0}cotanh({\epsilon\over {2kT}})$,
$\epsilon = {E_0\over N}$, ${E_0}^2 = {b{\mu}^2{c^4}N\over 4}$, and
$\overline{(\Delta E)^2} = N^{-1}{\overline{E}}^2 - {1 \over
4}N{\epsilon}^2 = N^{-1}{\overline{E}}^2 - {1 \over 4}b{\mu}^2c^4$.
Neglecting the ${1 \over N}$ term in this formula,
when $\overline{E}$ is fixed,
we obtain the expression for statistical fluctuations typical
of gravitating systems:
$\overline{{(\Delta E)^2}_{bh}} = -{1\over 4}b{\mu}^2c^4$.
The well known relation between the mass-energy fluctuations
and the behaviour of entropy near the state of thermal equilibrium,
$\overline{(\Delta E)^2} = - k\biggl({\partial^2 S \over \partial E^2}
\biggr)^{-1}$,
leads to the entropy of such a {\it truncated system}:
${\partial^2 S_{bh} \over \partial E^2} = 4kb^{-1}{\mu}^{-2}c^{-4}$.
Integrating this last equation gives the inverse temperature
$\beta_{bh} = 4b^{-1}\mu^{-2}c^{-4}{\overline{E}}$,
where an arbitrary
integration
constant is fixed to be zero by demanding that a very massive body
is also very cold~\cite{pom2}~\cite{gorpom}.
The entropy is given by the
`phenomenological' entropy formula of Bekenstein~\cite{bek}:
$S_{bh} = 2kb^{-1}{\mu}^{-2}{\overline{E}}^2 + const$.
The model calculation of Hawking~\cite{hawk} leads to a numerical
value of the constant $b$, $b = {1 \over 4{\pi}^2}$.
Quite independently of the actual value of the numerical constant $b$
the entropy $S_{bh}$ has a lower bound
$S_0 = 2kb^{-1}{\mu}^{-2}{E_0}^2 = k{N\over 2}$,
which depends only on $N$.

We have seen the emergence of the Bekenstein formula for the
black hole entropy from the hypothesis about the microscopic nature of
gravitational phenomena~\cite{pom1}~\cite{pom2}~\cite{pom5}~\cite{pom6}.
We shall propose to apply the same
argument to the whole Universe. It seems natural to consider $N_U\sim
10^{123}$ gravitational atoms~\cite{pom2}
and apply to them the same physical
argument. We obtain the simple estimate~\cite{pom2}
for cosmological constant
regarded as the zero-point vacuum energy density ${\lambda}\sim
{\mu^4\over N_U}$. It was shown~\cite{pom5} that a system of a large
number of gravitational atoms described in the framework
of the Atomic Theory of Gravitation (the new gravitational mechanics)
by a large $N\sim N_U$ matrix of operators~\cite{pom1}~\cite{pom2}
~\cite{pom5}~\cite{pom6} bears remarkable
similarity to the thermal Universe with a nonvanishing positive
cosmological constant.
%If the Bekenstein entropy were the whole thing as
%far as the thermal properties of gravitating masses are concerned, then
%the World would be always in a state of the lowest thermodynamic
%probability.
%%This conclusion would lead then to the statement that
%the behaviour of a visible Universe is determined by the condition
%that it is in a state of the lowest statistical weight.
%Considering an ensemble of such Universes, regarded as local
%thermal phenomena in a sense suggested in~\cite{pom2}~\cite{pom6},
%we would
%be persuaded to conclude that our Universe is the least probable one.
%The Universe must be regarded as a very typical one in the
%{\it statistical ensemble of Universes}, which is also the statement
%of the maximal thermodynamic probability $W$ of Boltzmann.
It should be noticed that the notion of a statistical ensemble
for the observable Universe is justified only after we identify
{\it atoms} whose existence is underlying the totality of phenomena.
The Gibbs-Jaynes principle of the maximum
of $H$ function~\cite{jay} is
applicable to a closed system once we postulate the general
{\it Atomic Hypothesis}~\cite{pom1}~\cite{pom1}~\cite{pom2}
~\cite{pom3}.
According to this hypothesis the totality of phenomena
should be derived from the properties of space-time-matter
atoms, which I prefer to call {\it gravitational atoms}
~\cite{pom1}~\cite{pom2}~\cite{pom3}~\cite{gorpom}.
The result described in this short note, which
was based
on the purely phenomenological considerations, was first derived in the
Spring of 1995 and published in~\cite{pom2}. I have
applied the simple
interpolation argument, originally due to Planck, to the problem of
mass-energy fluctuations of a gravitating
mass~\cite{pom2}~\cite{gorpom}. I have proposed that
the formalism of {\bf the new gravitational noncommutative mechanics}
~\cite{pom1}~\cite{pom2}~\cite{pom3}~\cite{pom5}~\cite{pom6}
be applied to gravitating particles (black holes)
and to the whole Universe. The point made very clearly
in~\cite{pom1} was that the microscopic theory of gravitation
requires new ideas of the kind Heisenberg has once introduced.
The idea of the {\em Second Heisenberg Algebra} was then
introduced~\cite{pom1}~\cite{pom3}~\cite{pom5}~\cite{pom6}.
%Hamilton's optical analogy should be taken seriously for both
%GRT and QM.
It is obvious to the present author that this idea is quite useful
once we realize that the perihelion precession of the Mercury, and
the {\em Complementarity Principle} for a gravitational mass and a
inertial mass aspect of a gravitating particle
~\cite{pom1}~\cite{pom3}~\cite{pom5}~\cite{pom6} allow for the
transition to new gravitational mechanics
in the same way the nonrelativistic Kepler
problem and the spectroscopic data has made it possible for Heisenberg
to propose the matrix quantum mechanics~\cite{pom1}.
The idea of using the GRT
Kepler problem to make a similar
transition to the new gravitational noncommutative mechanics was first
described by the present author~\cite{pom1}~\cite{pom3}
as early as in the Spring of 1995,
%(the submittal date of Ref ~\cite{pom1} was May 31, 1995),
and later during the USC Summer 1995 Institute~\cite{pom3}.
The idea of what
I have called the {\it Second Heisenberg Algebra}
and the {\it Second Period of Nature} was also described
in numerous lectures and talks
(CALTECH, February 1996; Tel-Aviv University, March 1996;
Ecole Polytechnique, Paris, October 1996).
The picture of a gravitating particle
(a black hole) which has emerged from my work is not unlike that of
a giant nucleus or a baryon\footnote{The large $N$ QCD description of
baryons and the Wigner (or the modern matrix models)
description of energy levels in nuclei share some common traits.}.
The calculation of specific properties of such a complex object
like a quantum black hole must be carried out in a sort of
large $N$ approximation in the new gravitational noncommutative
mechanics proposed
in~\cite{pom1}~\cite{pom3}~\cite{gorpom}~\cite{pom5}~\cite{pom6},
and it is under way.
\section*{Acknowledgments}
This research was partially supported by a NSF grant to USC.
I wish to thank organizers of the Eight Marcel Grossmann Meeting in
Jerusalem for inviting me to this Meeting, which, however, I
was unable to attend.

\end{document}